\documentclass[onecolumn,showpacs,showkeys,preprintnumbers]{revtex4}
\usepackage{amssymb}

\usepackage{amsmath}
\usepackage{graphicx}

\begin{document}

\title{Observational Constraints to Ricci Dark Energy Model by Using: SN, BAO, OHD, fgas Data Sets}
\author{Lixin Xu}
\email{lxxu@dlut.edu.cn}
\author{Yuting Wang}
\email{wangyuting0719@163.com}
\affiliation{School of Physics and Optoelectronic Technology,\\
Dalian University of Technology, Dalian, Liaoning 116024, P. R.
China}
\begin{abstract}
In this paper, we perform a global constraint on the Ricci dark
energy model with both the flat case and the non-flat case, using
the Markov Chain Monte Carlo (MCMC) method and the combined
observational data from the cluster X-ray gas mass fraction,
Supernovae of type Ia (397), baryon acoustic oscillations, current
Cosmic Microwave Background, and the observational Hubble function.
In the flat model, we obtain the best fit values of the parameters
in $1\sigma, 2\sigma$ regions: $\Omega_{m0}=0.2927^{+0.0420
+0.0542}_{-0.0323 -0.0388}$, $\alpha=0.3823^{+0.0331
+0.0415}_{-0.0418 -0.0541}$, $Age/Gyr=13.48^{+0.13 +0.17}_{-0.16
-0.21}$, $H_0=69.09^{+2.56 +3.09}_{-2.37 -3.39}$. In the non-flat
model, the best fit parameters are found in $1\sigma, 2\sigma$
regions:$\Omega_{m0}=0.3003^{+0.0367 +0.0429}_{-0.0371 -0.0423}$,
$\alpha=0.3845^{+0.0386 +0.0521}_{-0.0474 -0.0523}$,
$\Omega_k=0.0240^{+0.0109 +0.0133}_{-0.0130 -0.0153}$,
$Age/Gyr=12.54^{+0.51 +0.65}_{-0.37 -0.49}$, $H_0=72.89^{+3.31
+3.88}_{-3.05 -3.72}$. Compared to the constraint results in the
$\Lambda \textmd{CDM}$ model by using the same datasets, it is shown
that the current combined datasets prefer the $\Lambda \textmd{CDM}$
model to the Ricci dark energy model.
\end{abstract}

\maketitle

\address{Department of Physics, Dalian University of Technology,}
\section{Introduction}
The observation of the Supernovae of type Ia (SN Ia)
\cite{ref:Riess98,ref:Perlmuter99} provides the evidence that the
universe is undergoing accelerated expansion. Jointing current
Cosmic Microwave Background (CMB) anisotropy measurement from
Wilkinson Microwave Anisotropy Probe (WMAP)
\cite{ref:Spergel03,ref:Spergel06} and the data of the Large Scale
Structure (LSS) from SDSS \cite{ref:Tegmark1,ref:Tegmark2}, one
concludes that there exists an exotic energy component with negative
pressure, dubbed dark energy, whose density accounts for two-thirds
of the total energy density in the universe at present. The simplest
but most natural candidate of dark energy is the cosmological
constant ${\Lambda}$, with the constant equation of state
$w_{\Lambda}=-1$. In fact, the observational data is mostly
consistent with the predictions of the $\Lambda \textmd{CDM}$ model
\cite{ref:LCDM1,ref:sn192,ref:LCDM3}. However, it suffers the
so-called fine tuning and cosmic coincidence problems. To avoid
these problems, dynamic dark energy models are considered as an
alternative scenario, such as quintessence
\cite{ref:quintessence01,ref:quintessence02,ref:quintessence1,ref:quintessence2,ref:quintessence3,ref:quintessence4},
phantom \cite{ref:phantom}, quintom \cite{ref:quintom} and
holographic dark energy \cite{ref:holo1,ref:holo2} etc.

Although more and more dark energy models have been presented, the
nature of dark energy is still a conundrum. Under such
circumstances, the models which are built according to some
fundamental principle are more charming, such as holographic dark
energy model and agegraphic dark energy model
\cite{ref:age1,ref:age2}. The former one is on the basis of
holographic principle, and the latter one is derived from taking the
combination between the uncertainty relation in quantum mechanics
and general relativity into account. In this paper, we focus on the
holographic dark energy model, which is considered as a dynamic
vacuum energy and constructed by considering the holographic
principle and some features of quantum gravity theory. According to
the holographic principle, the number of degrees of freedom in a
bounded system should be finite and has relations with the area of
its boundary. By applying the principle to cosmology, one can obtain
the upper bound of the entropy contained in the universe. For a
system with size $L$ and UV cut-off $\Lambda$ without decaying into
a black hole, it is required that the total energy in a region of
size $L$ should not exceed the mass of a black hole of the same
size, thus $L^3\rho_\Lambda\leq LM_{pl}^2$. The largest $L$ allowed
is the one saturating this inequality, thus
\begin{eqnarray}
&&\rho_\Lambda=\frac{3c^2M_{pl}^2}{L^2},
\end{eqnarray}
where c is a numerical constant and $M_{pl}$ is the reduced Planck
Mass $M_{pl}\equiv1/\sqrt{8\pi G}$. It just means a duality between
UV cut-off and IR cut-off. The UV cut-off is related to the vacuum
energy, and IR cut-off is related to the large scale of the
universe, for example Hubble horizon, event horizon or particle
horizon as discussed by \cite{ref:holo1,ref:holo2}.

In the paper \cite{ref:holo2}, the author took the future event
horizon
\begin{eqnarray}
&&R_{eh}(a)=a\int_t^\infty\frac{dt'}{a(t')}=a\int_a^\infty\frac{da'}{Ha'^2}
\end{eqnarray}
as the IR cut-off $L$. Although this model is confronted with the
causality problem, as pointed out in \cite{ref:holo2}, it can reveal
the dynamic nature of the vacuum energy and provide a solution to
the fine tuning and cosmic coincidence problem. When $c\geq1$, $c=1$
and $c\leq1$, the holographic dark energy behaves like quintessence,
cosmological constant and phantom respectively. Therefore, in this
model, the value of parameter $c$ plays an important role in
determining the property of holographic dark energy. Then, a model
with holographic dark energy proportional to the Ricci scalar was
proposed by Gao, et. al. in \cite{ref:holo3}, called the Ricci dark
energy (RDE). In that paper \cite{ref:holo3}, it has shown that this
model can avoid the causality problem and naturally solve the
coincidence problem of dark energy after Ricci scalar is regarded as
the $IR$ cut-off $L^{-2}$:
\begin{eqnarray}
&&L^{-2}=R=-6(\dot{H}+2H^2+\frac{k}{a^2}).
\end{eqnarray}

An fascinating aspect in the study of dark energy is that the
cosmological parameters in a given model can be constrained by the
increasing observational data. It is found that the best way to
constrain is using the combination of a thorough observation. Now we
give a brief review of the previous works on the combined
observational constraints of the Ricci dark energy model. In Ref.
\cite{ref:MPLAXU}, we constrained the parameters $\Omega_{m0}$ and
$\alpha$ using 192 SN Ia data \cite{ref:sn192} from the ESSENCE
\cite{essence} and Gold sets
\cite{Riess:2004nr,Astier:2005qq,Riess:2006fw}, the CMB shift
parameter $R$ from three-year WMAP data \cite{ref:Wang06}, and the
BAO parameter $A$ from SDSS \cite{ref:Eisenstein05}, obtaining the
best-fittings: $\Omega_{m0}=0.34\pm0.04$ and $\alpha=0.38\pm0.03$.
Subsequently, the authors in Ref. \cite{ref:PRDzhang,ref:IRicci}
utilized the latest 307 Union SNIa data from the Supernova Cosmology
Project (SCP), the updated shift parameter $R$ from the five-year
WMAP data \cite{ref:Komatsu2008}, and the independent form of the
SDSS BAO parameter $A$ \cite{ref:Eisenstein05} to constrain the
parameters $\Omega_{m0}$ and $\alpha$, whose best-fit values are
given by $\Omega_{m0}=0.318_{-0.024}^{+0.026}$ and
$\alpha=0.359_{-0.025}^{+0.024}$ in \cite{ref:PRDzhang}. In Ref.
\cite{ref:GRicci}, the SDSS BAO parameter $A$ is replaced by the
measurement of $D_V(0.35)/D_V (0.2)$ from SDSS \cite{ref:Percival3}
to investigate the generalized Ricci dark energy model. Then Li, et.
al in \cite{ref:Li2} used the extended 397 SNIa data from the
Union+CFA3 sample \cite{ref:Condata}, only the values of
$[r_s(z_d)/D_V(0.2), r_s(z_d)/D_V(0.35)]$ in the measurement of BAO
\cite{ref:Percival2} and the Maximum likelihood values of
$[l_A(z_\ast), R(z_\ast), z_\ast]$ and their inverse covariance
matrix in the measurement of CMB \cite{ref:Komatsu2008}, getting the
best constraint results: $\Omega_{m0}=0.304, \alpha=0.363$.

In this paper, we will revisit the RDE model and make a thorough
investigation on this model with a completely consistent analysis of
the combined observations. Comparing with the previous works, we use
the more complete combinations of the observational datasets from
the X-ray gas mass fraction in clusters of galaxies (CBF)
\cite{ref:07060033}, 397 SN Ia \cite{ref:Condata} data, the BAO
measurement on the values of $[r_s(z_d)/D_V(0.2),
r_s(z_d)/D_V(0.35)]$ and their inverse covariance matrix in
\cite{ref:Percival2}, the CMB observation \cite{ref:Komatsu2008} on
the Maximum likelihood values of $[l_A(z_\ast), R(z_\ast), z_\ast]$
and their inverse covariance matrix and the observational Hubble
data at fifteen different redshifts, including the three more
observational data $H(z=0.24)=79.69\pm2.32, H(z=0.34)=83.8\pm2.96,$
and $H(z=0.43)=86.45\pm3.27$ in \cite{ref:0807} and the
observational data \cite{ref:0905,ref:0907}. We carry out the global
fitting on the RDE model using the Markov Chain Monte Carlo (MCMC)
method. In addition, in this paper we do not only perform the
constraint on the parameters in the flat RDE model, but also in the
non-flat RDE model, comparing with the constraint results in the
standard concordance model by using the same combined datasets.

The paper is organized as follows. In next section, we briefly
review the RDE model. In section III, we describe the method and
data. After we perform the cosmic observation constraint, the
results on the determination of the cosmological parameters are
presented. The last section is the conclusion.

\section{Review the Ricci Dark Energy Model}

In this section, we give a brief review on the general formula in
the RDE model. With a Friedmann-Robertson-Walker (FRW) metric
\begin{eqnarray}
&&ds^2=-dt^2+a^2(t)[\frac{dr^2}{1-kr^2}+r^2(d{\theta}^2+\sin^2{\theta}d{\phi}^2)],
\end{eqnarray}
the Einstein field equation can be written as
\begin{eqnarray}
&&H^2=\frac{1}{3M_{pl}^2}(\rho_m+\rho_R)-\frac{k}{a^2},
\end{eqnarray}
where the parameter $k$ denotes the curvature of space $k=1,0,-1$
for closed, flat and open geometries, respectively, $H$ is the
Hubble function, and ${\rho}$ is the energy density of a general
piece of matter, and their subscripts $m$ and $R$ respectively
correspond to matter component, including the cold dark matter
$\rho_{cdm}$ and baryon matter $\rho_b$, and RDE.

As suggested by Gao et. al., the energy density of RDE is
proportional to the Ricci scalar, thus whose energy density is given
as
\begin{eqnarray}
&&\rho_R=3\alpha M_{pl}^2(\dot{H}+2H^2+\frac{k}{a^2})\propto R,
\end{eqnarray}
where $\alpha$ is the dimensionless parameter in RDE model, which
can be determined though cosmic observation constraints. After
changing the variable from the cosmic time $t$ to $x=\ln a$, we can
rewritten the Friedmann Eq. (5) as
\begin{eqnarray}
&&H^2=\frac{1}{3M_{pl}^2}\rho_{m0}e^{-3x}+(\alpha-1)ke^{-2x}+\alpha(\frac{1}{2}\frac{dH^2}{dx}+2H^2).
\end{eqnarray}
With the help of the definitions as follows:
\begin{eqnarray}
&&E=
\frac{H}{H_0},\Omega_{m0}=\frac{\rho_{m0}}{3M_{pl}^2H_0^2},\Omega_k=-\frac{k}{H_0^2},
\end{eqnarray}
the Eq. (7) can be ulteriorly rewritten as
\begin{eqnarray}
&&E^2=(1-\alpha)\Omega_ke^{-2x}+\Omega_{m0}e^{-3x}+\alpha(\frac{1}{2}\frac{dE^2}{dx}+2E^2).
\end{eqnarray}
Solving this first order differential equation about $E^2$, we can
obtain
\begin{eqnarray}
E^2&&{=}\Omega_ke^{-2x}+\Omega_{m0}e^{-3x}+\frac{\alpha}{2-\alpha}\Omega_{m0}e^{-3x}+f_0e^{-(4-\frac{2}{\alpha})x} \nonumber\\
&&{=}\Omega_ke^{-2x}+\Omega_{m0}e^{-3x}+\Omega_{R}(x),
\end{eqnarray}
where $f_0$ is the integral constant and can be derived by the
initial condition $E(x=0)=1$, i.e.
$\Omega_k+\Omega_{m0}+\Omega_{R0}=1$ is used, which is
$f_0=1-\Omega_k-\frac{2}{2-\alpha}\Omega_{m0}$, and $\Omega_R$ is
the definition of the dimensionless RDE density, with the expression
being
\begin{eqnarray}
&&\Omega_R(x)=\frac{\alpha}{2-\alpha}\Omega_{m0}e^{-3x}+(1-\Omega_k-\frac{2}{2-\alpha}\Omega_{m0})e^{-(4-\frac{2}{\alpha})x}.
\end{eqnarray}

\section{Method and results}

\begin{table}
\begin{center}
\begin{tabular}{cc|   cc   cc|   cc   cc}
\hline\hline model parameters & & flat RDE  &  &flat $\Lambda
\textmd{CDM}$ & & not-flat RDE & &non-flat $\Lambda \textmd{CDM}$ &
\\ \hline
$\Omega_{m0}$    && $0.2927^{+0.0420 +0.0542}_{-0.0323 -0.0388}$ &
                     & $0.2778^{+0.0268 +0.0396}_{-0.0323 -0.0379}$ &
                     & $0.3003^{+0.0367 +0.0429}_{-0.0371 -0.0423}$ &
                     & $0.2730^{+0.0352 +0.0424}_{-0.0303 -0.0330}$ & \\
$\alpha$         && $0.3823^{+0.0331 +0.0415}_{-0.0418 -0.0541}$ &
                     & - &
                     & $0.3845^{+0.0386 +0.0521}_{-0.0474 -0.0523}$ &
                     & - & \\
$\Omega_{k}$        && - &
                     & - &
                     & $0.0240^{+0.0109 +0.0133}_{-0.0130 -0.0153}$ &
                     & $-0.0010^{+0.0106 +0.0135}_{-0.0105 -0.0139}$ & \\
$Age/Gyr$           && $13.48^{+0.13 +0.17}_{-0.16 -0.21}$ &
                     & $13.71^{+0.11 +0.13}_{-0.13 -0.16}$ &
                     & $12.54^{+0.51 +0.65}_{-0.37 -0.49}$ &
                     & $13.75^{+0.51 +0.65}_{-0.48 -0.59}$ & \\
$H_0$               && $69.09^{+2.56 +3.09}_{-2.37 -3.39}$ &
                     & $70.23^{+2.56 +3.19}_{-1.91 -2.70}$ &
                     & $72.89^{+3.31 +3.88}_{-3.05 -3.72}$ &
                     & $70.38^{+2.69 +3.18}_{-2.70 -3.57}$ & \\
 \hline $\chi^{2}_{min}$  && 562.543 &
                     & 521.669  &
                     & 539.763  &
                     & 521.557  &   \\
\hline\hline
\end{tabular}
\caption{The data fitting results of the parameters with $1\sigma$,
$2\sigma$ regions in RDE model and the $\Lambda\textmd{CDM}$ model,
where the combined observational data from CBF, SN, BAO and CMB and
OHD are used.}\label{tab:results}
\end{center}
\end{table}

\begin{figure}
  \includegraphics[width=530pt]{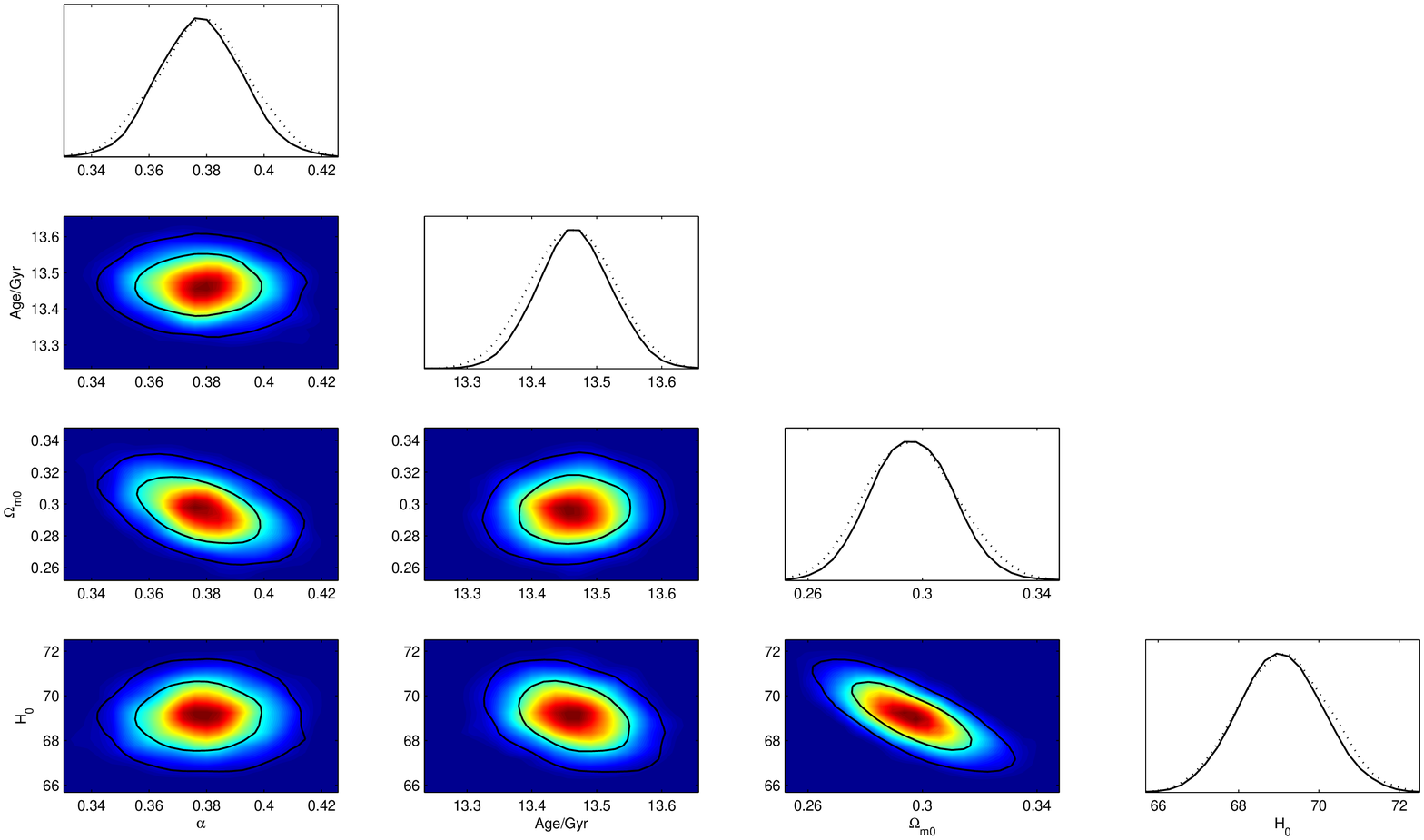}
  \caption{1-D constraints on individual parameters ($\alpha, Age/Gyr, \Omega_{m0},
  H_0$) and 2-D contours on these parameters with $1\sigma, 2\sigma$ errors between each other using the combination of the observational
  data from CBF, SN, BAO, CMB and OHD in the flat RDE model. Dotted lines in the 1-D plots
  show the mean likelihood of the samples and the solid lines are marginalized
  probabilities for the parameters in the flat RDE model \cite{ref:MCMC}.
 }\label{fig:flat}
\end{figure}

\begin{figure}
  \includegraphics[width=530pt]{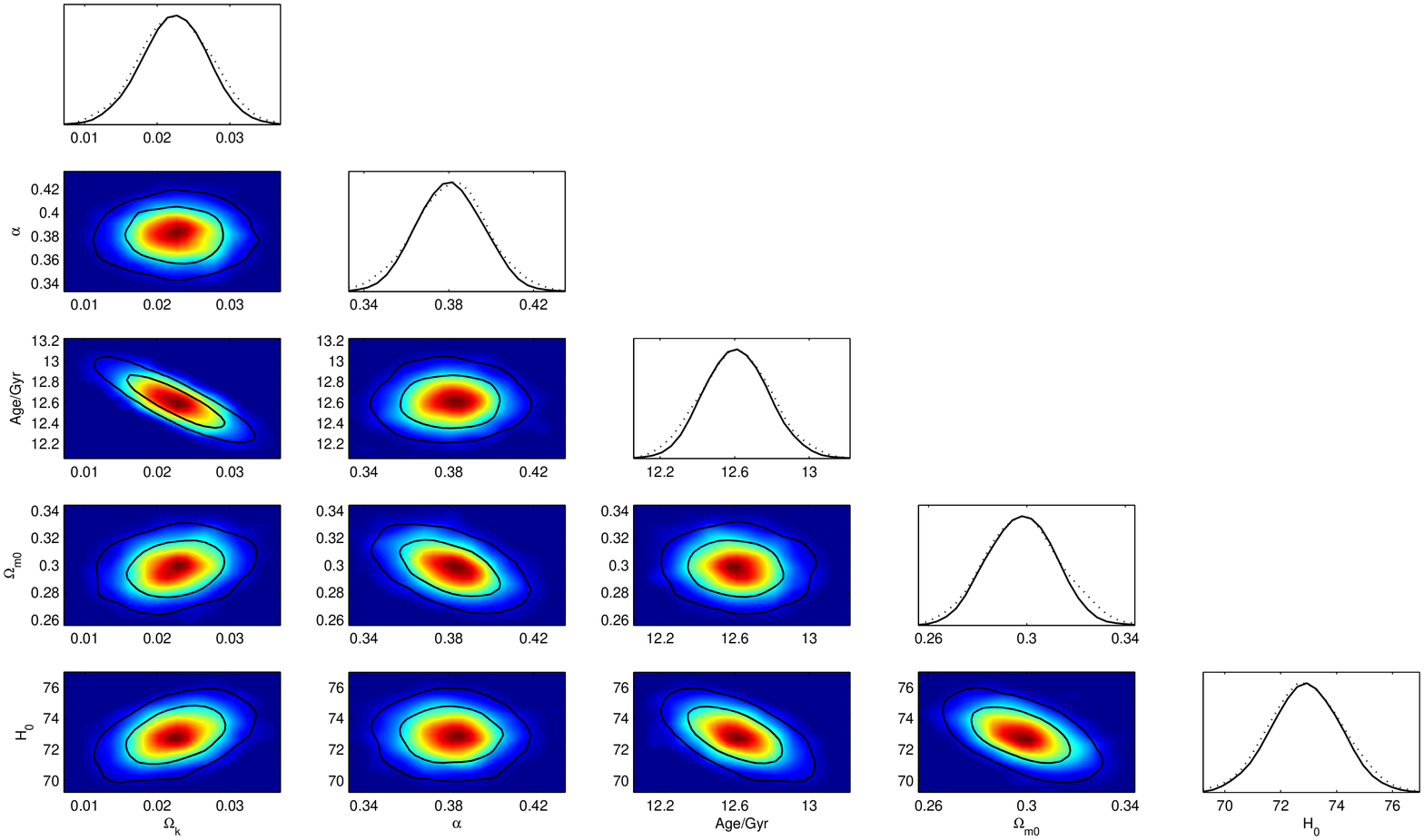}
  \caption{1-D constraints on individual parameters ($\Omega_k, \alpha, Age/Gyr, \Omega_{m0},
  H_0$) and 2-D contours on these parameters with $1\sigma, 2\sigma$ errors between each other using the combination of the observational
  data from CBF, SN, BAO, CMB and OHD in the non-flat RDE model. Dotted lines in the 1-D plots
  show the mean likelihood of the samples and the solid lines are marginalized
  probabilities for the parameters in the non-flat RDE model \cite{ref:MCMC}.
 }\label{fig:nonflat}
\end{figure}
In our analysis, we perform a global fitting on determining the
cosmological parameters using the Markov Chain Monte Carlo (MCMC)
method. The MCMC method is based on the publicly available {\bf
CosmoMC} package \cite{ref:MCMC} and the {\bf modified CosmoMC}
package \cite{ref:0409574,ref:07060033,ref:modifiedMCMC}, including
the X-ray cluster gas mass fraction. For our models we has modified
these packages to add the new parameter $\alpha$ with a prior
$\alpha\in[0.1,0.8]$. Besides the parameter $\alpha$, the following
basic cosmological parameters ($\Omega_bh^2, \Omega_ch^2,
\Omega_k$)are also varying with top-hat priors: the physical baryon
density $\Omega_{b}h^2\in[0.005,0.1]$, the physical dark matter
energy density $\Omega_{c}h^2\in[0.01,0.99]$, and the density of
space curvature $\Omega_k\in[-0.1,0.1]$. The cosmological parameter
$\Omega_{m0}$ can be derived from the basic parameters above. We use
a top-hat prior of the cosmic age i.e.$10 Gyr<t_0<20 Gyr$ and impose
a weak Gaussian prior on the physical baryon density
$\Omega_{b}h^2=0.022\pm0.002$ from Big Bang nucleosynthesis
\cite{ref:bbn}. Also, in the data fitting process the seven
parameters ($K, \eta, \gamma, b_0, \alpha_b, s_0, \alpha_s$)
included in the X-ray gas mass fraction $f_{gas}$ are treated as
free parameters. As a byproduct the best fitting values of these
parameters are obtained. And, these values can also be taken
accounted as a check of data fitting.

In our calculations, we have taken the total likelihood function
$L\propto e^{-\chi^2/2}$ to be the products of the separate
likelihoods of CBF, SN, BAO, CMB and OHD. Then we get $\chi^2$ is
\begin{eqnarray}
\chi^2=\chi^2_{CBF}+\chi^2_{SN}+\chi^2_{BAO}+\chi^2_{CMB}+\chi^2_{OHD},
\end{eqnarray}
where the separate likelihoods of CBF, SN, BAO, CMB and OHD and the
current observational datasets used in this paper are shown in the
Appendix \ref{app}.

The results on the best values of the cosmological parameters with
$1\sigma, 2\sigma$ errors in the RDE model and the $\Lambda
\textmd{CDM}$ model are listed in Table \ref{tab:results}. In the
Fig. \ref{fig:flat}, we show one dimensional probability
distribution of each parameter and two dimensional plots for
parameters between each other in the flat RDE model. The
corresponding plots in the non-flat RDE model are presented in Fig.
\ref{fig:nonflat}. From Figs. \ref{fig:flat} and \ref{fig:nonflat},
it is seen that the cosmological parameters in the two cases of RDE
model are well determined in $1\sigma$ and $2\sigma$ regions.
Comparing the flat RDE model with the non-flat RDE model, we can
find the difference of $\chi^2_{min}$ is obvious. The non-flat RDE
model with a smaller $\chi^2_{min}=539.763$ is favored over the flat
case with $\chi^2_{min}=562.543$. However, compared to the $\Lambda
\textmd{CDM}$ model, it is found the $\Lambda \textmd{CDM}$ models
with smaller values $\chi^2_{min}=521.669$ (flat case) and
$\chi^2_{min}=521.557$ (non-flat case) are better fit to the current
combined data than the RDE model. What is more, we obtain the
best-fit values of the parameters in $f_{gas}$: $K=0.9665,
\eta=0.2058, \gamma=1.0866, b_0=0.7073, \alpha_b=-0.0540,
s_0=0.1654, \alpha_s=0.1591$ in the flat case and $K=0.9871,
\eta=0.2114, \gamma=1.0507, b_0=0.7749, \alpha_b=-0.0950,
s_0=0.1741, \alpha_s=0.0194$ in the non-flat case.

\begin{figure}
  \includegraphics[width=350pt]{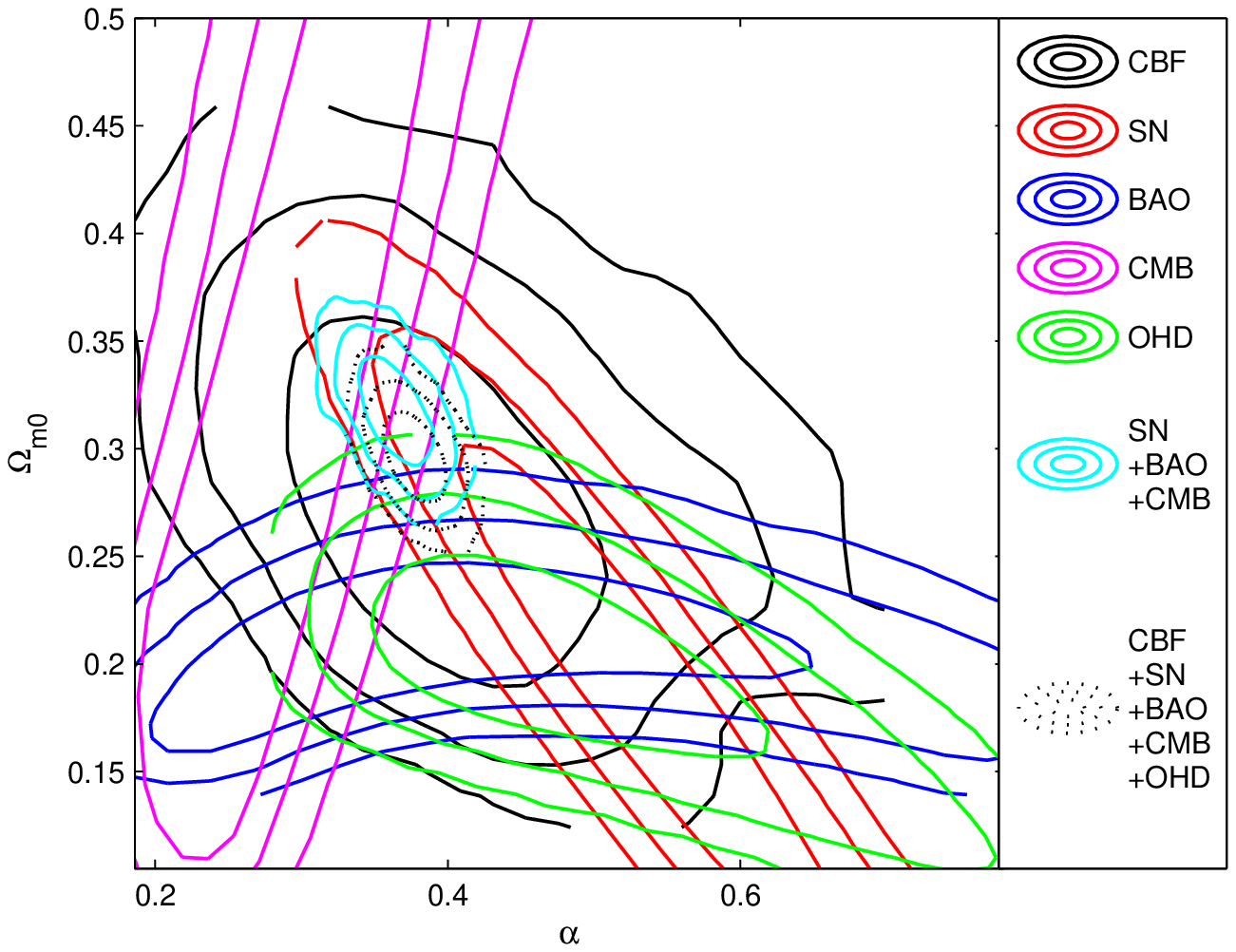}
  \caption{The 2-D contours on the parameters ($\alpha$, $\Omega_{m0}$) with $1\sigma$,
  $2\sigma$ and $3\sigma$ confidence in the flat RDE model for CBF only (black solid line), SN only (red line),
  BAO only (blue line), CMB only (magenta line), OHD only (green line). The joint constraint from
  SN, BAO and CMB is shown in cyan line and the fully combined
  constraint from CBF, SN, BAO, CMB, and OHD is shown in black dotted
  line.
 }\label{fig:Omegaalpha}
\end{figure}

From Fig. \ref{fig:Omegaalpha}, we can see the constraints on the
parameters ($\alpha$, $\Omega_{m0}$) in the flat RDE model by using
the independent dataset from CBF (black solid), SN (red), BAO
(blue), CMB (magenta) and OHD (green) and the combined datesets
(black dotted). It is found all the constraint results from the
independent dataset are consistent. It is obvious that the space of
the parameters($\alpha$, $\Omega_{m0}$) has been reduced by using
the three combined datasets from SN, BAO and CMB (cyan). It is
indicated that the tight constraint has been obtained by the three
combined datasets. Combined with the additional datasets from CBF
and OHD, it is seen that the best fit values of the parameters are
changed. For the parameters ($\alpha$, $\Omega_{m0}$), the inclusion
of the datasets from CBF and OHD changes their best fit values from
($0.3630, 0.3192$) to ($0.3823,0.2927$). In addition, as shown in
Fig. \ref{fig:Omegaalpha}, though the degeneracy with the additional
datasets from CBF and OHD is not obviously improved, the curves of
the 2-D contour plot from the five combined datasets become smoother
than that from the three combined datasets.

Next, we investigate the models according to the objective
Information Criterion (IC) including the Akaike Information
Criterion (AIC) and Bayesian Information Criterion (BIC). Here, we
give a brief introduction, for the details please see in Ref.
\cite{ref:AIC,ref:AICBIC,ref:AICBiesiada,ref:BIC,ref:xuCPL}. The AIC
was based on the Kullback-Leibler information entropy and derived by
H. Akaike. It takes the form
\begin{equation}
AIC=-2\ln \mathcal{L}(\hat{\theta}|data)_{max}+2K,\label{eq:AIC}
\end{equation}
where, $\mathcal{L}_{max}$ is the highest likelihood in the model
with the best fit parameters $\hat{\theta}$, $K$ is the number of
parameters in the model. The first term of Eq. (\ref{eq:AIC})
measures the goodness of model fit, and the second one interprets
model complexity. The BIC is similar as AIC, but the second term is
different. It was derived by G. Schwarz and is given by the form
\begin{equation}
BIC=-2\ln \mathcal{L}(\hat{\theta}|data)_{max}+K\ln n,\label{eq:BIC}
\end{equation}
where, $n$ in the different term is the number of data points in the
datasets. In the above two cases, the term $-2\ln
\mathcal{L}(\hat{\theta}|data)_{max}$ is often called
$\chi^2_{min}$, though it is also generalized to non-Gaussian
distributions.

Now, the problem is how to evaluate which model is the better one.
It is the issue of strength of evidence. We take AIC case as a
example. And, BIC is the same as that. Comparing the AIC values of
several models, the minimum one is considered as the best value and
denoted by $AIC_{min}=min\{AIC_i,i=1...N\}$, where $i=1...N$ is a
set of alternative candidate models. The difference of AIC for other
alternative model $i$ with respect to the besic model is expressed
as $\Delta AIC_i$=$AIC_i$-$AIC_{min}$. The rules of judgement of AIC
model selection are that when $\Delta AIC_i$ is in the range $0-2$,
it means the model $i$ has almost the same supports from the data as
the best one; for the range $2-4$, this support is considerably less
and with $\Delta AIC_i>10$ model $i$ is practically irrelevant
\cite{ref:AICBiesiada}. For BIC, the rules read that a $\Delta$BIC
of more than 2 (or 6) relative to the best one is considered
"unsupported" (or "strongly unsupported") from observational data
\cite{ref:AIC}.

\begin{table}[ht]
\begin{center}
\begin{tabular}{cc|    cc    cc|  cc   cc|   cc   cc}
\hline\hline Model & & Parameters & & $\chi^2_{min}$& & AIC  &  &
$\Delta$AIC & & BIC & & $\Delta$BIC &
\\ \hline
flat $\Lambda$CDM && 9 &        & 521.669&
                   & 539.669 &  & 0 &
                   & 576.83 &   & 0 &\\
non-flat $\Lambda$CDM && 10 &        & 521.557&
                   & 541.557 &  & 1.888 &
                   & 582.848 &   & 6.017 &\\
flat RDE           && 10 &        & 562.543&
                   & 582.543 &  & 42.874&
                   & 623.834 &   & 47.003 &\\
non-flat RDE      && 11 &        & 539.763&
                   & 561.763 &  & 22.094 &
                   & 607.183 &   &30.352&\\
\hline\hline
\end{tabular}
\caption{The results of $\chi^2_{min}$, AIC and BIC in the
$\Lambda$CDM model and RDE model and the differences of AIC and BIC
with respect to the flat $\Lambda$CDM model are
listed.}\label{tab:Information-criteria}
\end{center}
\end{table}

The values of AIC and BIC in the $\Lambda$CDM model and RDE model
and the differences of AIC and BIC with respect to the flat
$\Lambda$CDM model are listed in Table
\ref{tab:Information-criteria}. In the table, it reads the flat
$\Lambda$CDM model are favored by both AIC and BIC. According to the
rules of judgement of AIC model selection above, it is seen that the
non-flat $\Lambda$CDM model is supported considerably and the RDE
models are disfavored. Since BIC imposes a stricter penalty against
introducing extra parameters than AIC, as shown from Eq.
(\ref{eq:BIC}), all the models except the flat $\Lambda$CDM model
are disfavored by BIC.

\section{Conclusion}
In summary, in this paper we have performed a global fitting on the
cosmological parameters in both the flat RDE model and the non-flat
RDE model by using a completely consistent analysis of the X-ray gas
mass fraction observation, type Ia supernovae data from the
combination of CfA3 samples and the Union set, baryon acoustic
oscillations data from SDSS, the measurement data on current Cosmic
Microwave Background from the five-year WMAP observations and the
observational Hubble data. The constraint on the parameters in the
$\Lambda \textmd{CDM}$ model are performed by using the same
combined datasets. The constraint results are shown in Table
\ref{tab:results} for the flat RDE model $\chi^2_{min}=562.543$, for
the non-flat RDE model $\chi^2_{min}=539.763$, for the flat $\Lambda
\textmd{CDM}$ model $\chi^2_{min}=521.669$, and for the non-flat
$\Lambda \textmd{CDM}$ model $\chi^2_{min}=521.557$. From Fig.
\ref{fig:Omegaalpha}, it is shown that the best fit values of the
parameters ($\alpha,\Omega_{m0}$) are changed by the additional
datasets from CBF and OHD, though the additional data have a minor
effect on the confidences of the parameters. It is indicated that
more accurate data is anticipated to provide the more stringent
constraint on the parameters in RDE model. Comparing the RDE model
with the $\Lambda \textmd{CDM}$ model, we can find that the current
combined datasets do favor the $\Lambda \textmd{CDM}$ model over the
RDE model. According to AIC and BIC, we find that the flat
$\Lambda$CDM model is much supported by the current data. The RDE
model is disfavored by AIC and BIC.
\section*{Acknowledgments}
The data fitting is based on the publicly available {\bf CosmoMC}
package a Markov Chain Monte Carlo (MCMC) code. This work is
supported by the National Natural Science Foundation of China (Grant
No 10703001), and Specialized Research Fund for the Doctoral Program
of Higher Education (Grant No 20070141034).

\appendix

\section{Cosmological Constraints Methods}\label{app}
\subsection{The X-ray gas mass fraction constraints}
According to the X-ray cluster gas mass fraction observation, the
baryon mass fraction in clusters of galaxies (CBF) can be utilized
to constrain cosmological parameters. The X-ray gas mass fraction,
$f_{gas}$, is defined as the ratio of the X-ray gas mass to the
total mass of a cluster, which is approximately independent on the
redshift for the hot $(kT\gtrsim5keV)$, dynamically relaxed clusters
at the radii larger than the innermost core $r_{2500}$. The X-ray
gas mass fraction, $f_{gas}$, can be derived from the observed X-ray
surface brightness profile and the deprojected temperature profile
of X-ray gas under the assumptions of spherical symmetry and
hydrostatic equilibrium. Basing on these assumptions above, Allen et
al. \cite{ref:07060033} selected 42 hot $(kT\gtrsim5keV)$, X-ray
luminous, dynamically relaxed clusters for $f_{gas}$ measurements.
The stringent restriction to the selected sample can not only reduce
maximally the effect of the systematic scatter in $f_{gas}$ data,
but also ensure that the $f_{gas}$ data is independent on
temperature. In the framework of the $\Lambda \textmd{CDM}$
reference cosmology, the X-ray gas mass fraction is presented as
\cite{ref:07060033}
\begin{eqnarray}
&&f_{gas}^{\Lambda \textmd{CDM}}(z)=\frac{K A \gamma
b(z)}{1+s(z)}\left(\frac{\Omega_b}{\Omega_m}\right)
\left[\frac{d_A^{\Lambda \textmd{CDM}}(z)}{d_A(z)}\right]^{1.5},\ \
\ \ \label{eq:f_g}
\end{eqnarray}
where $A$ is the angular correction factor, which is caused by the
change in angle for the current test model $\theta_{2500}$ in
comparison with that of the reference cosmology
$\theta_{2500}^{\Lambda CDM}$:
\begin{eqnarray}
&&A=\left(\frac{\theta_{2500}^{\Lambda
\textmd{CDM}}}{\theta_{2500}}\right)^\eta \approx
\left(\frac{H(z)d_A(z)}{[H(z)d_A(z)]^{\Lambda
\textmd{CDM}}}\right)^\eta,
\end{eqnarray}
here, the index $\eta$ is the slope of the $f_{gas}(r/r_{2500})$
data within the radius $r_{2500}$, with the best-fit average value
$\eta=0.214\pm0.022$ \cite{ref:07060033}. And the angular diameter
distance is given by
\begin{eqnarray}
&&d_A(z)=\frac{c}{(1+z)\sqrt{|\Omega_k|}}\mathrm{sinn}[\sqrt{|\Omega_k|}\int_0^z\frac{dz'}{H(z')}],
\end{eqnarray}
where $\mathrm{sinnn}(\sqrt{|\Omega_k|}x)$ respectively denotes
$\sin(\sqrt{|\Omega_k|}x)$, $\sqrt{|\Omega_k|}x$,
$\sinh(\sqrt{|\Omega_k|}x)$ for $\Omega_k<0$, $\Omega_k=0$ and
$\Omega_k>0$.

In equation (\ref{eq:f_g}), the parameter $\gamma$ denotes
permissible departures from the assumption of hydrostatic
equilibrium, due to non-thermal pressure support; the bias factor
$b(z)= b_0(1+\alpha_b z)$ accounts for uncertainties in the cluster
depletion factor; $s(z)=s_0(1 +\alpha_s z)$ accounts for
uncertainties of the baryonic mass fraction in stars and a Gaussian
prior for $s_0$ is employed, with $s_0=(0.16\pm0.05)h_{70}^{0.5}$
\cite{ref:07060033}; the factor $K$ is used to describe the combined
effects of the residual uncertainties, such as the instrumental
calibration and certain X-ray modelling issues, and a Gaussian prior
for the 'calibration' factor is considered by $K=1.0\pm0.1$
\cite{ref:07060033};

Following the method in Ref. \cite{ref:CBFchi21,ref:07060033} and
adopting the updated 42 observational $f_{gas}$ data in Ref.
\cite{ref:07060033}, the best fit values of the model parameters for
the X-ray gas mass fraction analysis are determined by minimizing,
\begin{eqnarray}
&&\chi^2_{CBF}=\sum_i^N\frac{[f_{gas}^{\Lambda
\textmd{CDM}}(z_i)-f_{gas}(z_i)]^2}{\sigma_{f_{gas}}^2(z_i)},
\end{eqnarray}
where $\sigma_{f_{gas}}(z_i)$ is the statistical uncertainties
(Table 3 of \cite{ref:07060033}). As pointed out in
\cite{ref:07060033}, the acquiescent systematic uncertainties have
been considered according to the parameters i.e. $\eta, b(z), s(z)$
and $K$.

\subsection{Type Ia Supernovae constraints}

We use the 397 SN Ia Constitution dataset \cite{ref:Condata}. The 90
SN Ia from CfA3 sample with low redshifts are added to 307 SN Ia
Union sample \cite{ref:Kowalski}. The CfA3 sample increases the
number of the nearby SN Ia and reduces the statistical
uncertainties. In our analysis we use the Constitution datasets used
SALT fitter \cite{ref:SALT} to fit the SN Ia light curves, where the
intrinsic uncertainty of 0.138 mag for each CfA3 SNIa, the peculiar
velocity uncertainty of 400km/s, and the redshift uncertainty of
0.001 have been considered to realize a more cautious assumption
that there is the same Hubble residual uncertainty between the CfA3
SN and the nearby Union SN \cite{ref:Condata}. Following
\cite{ref:smallomega,ref:POLARSKI}, one can obtain the corresponding
constraints by fitting the distance modulus $\mu(z)$ as
\begin{equation}
\mu_{th}(z)=5\log_{10}[D_{L}(z)]+\mu_{0}.
\end{equation}
In this expression $D_{L}(z)$ is the Hubble-free luminosity distance
$H_0 d_L(z)/c=H_0 d_A(z)(1+z)^2/c$, with $H_0$ the Hubble constant,
defined through the re-normalized quantity $h$ as $H_0=100 h~{\rm km
~s}^{-1} {\rm Mpc}^{-1}$, and $\mu_0\equiv42.38-5\log_{10}h$.

 Additionally, the observed distance moduli $\mu_{obs}(z_i)$ of SN
Ia at $z_i$ is
\begin{equation}
\mu_{obs}(z_i) = m_{obs}(z_i)-M,
\end{equation}
where $M$ is their absolute magnitudes.

For the SN Ia dataset, the best fit values of the parameters $p_s$
  can be determined by a likelihood analysis, based on
the calculation of
\begin{eqnarray}
\chi^2(p_s,M^{\prime})\equiv \sum_{SN}\frac{\left\{
\mu_{obs}(z_i)-\mu_{th}(p_s,z_i)\right\}^2} {\sigma_i^2} \ \ \ \ \ \ \ \ \ \  \ \ \ \ \ \ \nonumber\\
=\sum_{SN}\frac{\left\{ 5 \log_{10}[D_L(p_s,z_i)] - m_{obs}(z_i) +
M^{\prime} \right\}^2} {\sigma_i^2}, \ \ \ \ \label{eq:chi2}
\end{eqnarray}
where $M^{\prime}\equiv\mu_0+M$ is a nuisance parameter which
includes the absolute magnitude and the parameter $h$. The nuisance
  parameter $M^{\prime}$ can be marginalized over
analytically \cite{ref:SNchi2} as
\begin{equation}
\bar{\chi}^2(p_s) = -2 \ln \int_{-\infty}^{+\infty}\exp \left[
-\frac{1}{2} \chi^2(p_s,M^{\prime}) \right] dM^{\prime},\nonumber
\label{eq:chi2marg}
\end{equation}
resulting to
\begin{equation}
\bar{\chi}^2 =  A - \frac{B^2}{C} + \ln \left( \frac{C}{2\pi}\right)
, \label{eq:chi2mar}
\end{equation}
with
\begin{eqnarray}
&&A=\sum_{SN} \frac {\left\{5\log_{10}
[D_L(p_s,z_i)]-m_{obs}(z_i)\right\}^2}{\sigma_i^2},\nonumber\\
&& B=\sum_{SN} \frac {5
\log_{10}[D_L(p_s,z_i)]-m_{obs}(z_i)}{\sigma_i^2},\nonumber
\\
&& C=\sum_{SN} \frac {1}{\sigma_i^2}\nonumber.
\end{eqnarray}
Relation (\ref{eq:chi2}) has a minimum at the nuisance parameter
value $M^{\prime}=B/C$, which contains information of the values of
$h$ and $M$. Therefore, one can extract the values of $h$ and $M$
provided the knowledge of one of them. Finally, note that the
expression
\begin{equation}
\chi^2_{SN}(p_s,B/C)=A-(B^2/C),\label{eq:chi2SN}\nonumber
\end{equation}
which coincides to (\ref{eq:chi2mar}) up to a constant, is often
used in the likelihood analysis
\cite{ref:smallomega,ref:JCAPXU,ref:SNchi2}, and thus in this case
the results will not be affected by a flat $M^{\prime}$
distribution.

\subsection{Baryon Acoustic Oscillation constraints}

The Baryon Acoustic Oscillations are detected in the clustering of
the combined 2dFGRS and SDSS main galaxy samples, and measure the
distance-redshift relation at $z = 0.2$. Additionally, Baryon
Acoustic Oscillations in the clustering of the SDSS luminous red
galaxies measure the distance-redshift relation at $z = 0.35$. The
observed scale of the BAO calculated from these samples, as well as
from the combined sample, are jointly analyzed using estimates of
the correlated errors to constrain the form of the distance measure
$D_V(z)$ \cite{ref:Okumura2007,ref:Percival2,ref:Eisenstein05}
\begin{equation}
D_V(z)=c\left(\frac{z}{\Omega_k
H(z)}\mathrm{sinn}^2[\sqrt{|\Omega_k|}\int_0^z\frac{dz'}{H(z')}]\right)^{1/3}.
\label{eq:DV}
\end{equation}
The peak positions of the BAO depend on the ratio of $D_V(z)$ to the
sound horizon size at the drag epoch (where baryons were released
from photons) $z_d$, which can be obtained by using a fitting
formula \cite{ref:Eisenstein}:
\begin{eqnarray}
&&z_d=\frac{1291(\Omega_mh^2)^{0.251}}{1+0.659(\Omega_mh^2)^{0.828}}[1+b_1(\Omega_bh^2)^{b_2}],
\end{eqnarray}
with
\begin{eqnarray}
&&b_1=0.313(\Omega_mh^2)^{-0.419}[1+0.607(\Omega_mh^2)^{0.674}], \\
&&b_2=0.238(\Omega_mh^2)^{0.223}.
\end{eqnarray}
In this paper, we use the data of $r_s(z_d)/D_V(z)$, which are
listed in Table \ref{baodata}, where $r_s(z)$ is the comoving sound
horizon size
\begin{eqnarray}
r_s(z)&&{=}c\int_0^t\frac{c_sdt}{a}=c\int_0^a\frac{c_sda}{a^2H}=c\int_z^\infty
dz\frac{c_s}{H(z)} \nonumber\\
&&{=}\frac{c}{\sqrt{3}}\int_0^{1/(1+z)}\frac{da}{a^2H(a)\sqrt{1+(3\Omega_b/(4\Omega_\gamma)a)}},
\end{eqnarray}
where $c_s$ is the sound speed of the photon$-$baryon fluid
\cite{ref:Hu1, ref:Hu2, ref:Caldwell}:
\begin{eqnarray}
&&c_s^{-2}=3+\frac{9}{4}\times\frac{\rho_b(z)}{\rho_\gamma(z)}=3+\frac{9}{4}\times(\frac{\Omega_b}{\Omega_\gamma})a,
\end{eqnarray}
and here $\Omega_\gamma=2.469\times10^{-5}h^{-2}$ for
$T_{CMB}=2.725K$.

\begin{table}[htbp]
\begin{center}
\begin{tabular}{c|l}
\hline\hline
 $z$ &\ $r_s(z_d)/D_V(z)$  \\ \hline
 $0.2$ &\ $0.1905\pm0.0061$  \\ \hline
 $0.35$  &\ $0.1097\pm0.0036$  \\
\hline
\end{tabular}
\end{center}
\caption{\label{baodata} The observational $r_s(z_d)/D_V(z)$
data~\cite{ref:Percival2}.}
\end{table}
Using the data of BAO in Table \ref{baodata} and the inverse
covariance matrix $V^{-1}$ in \cite{ref:Percival2}:

\begin{eqnarray}
&&V^{-1}= \left(
\begin{array}{cc}
 30124.1 & -17226.9 \\
 -17226.9 & 86976.6
\end{array}
\right),
\end{eqnarray}

Thus, the $\chi^2_{BAO}(p_s)$ is given as
\begin{equation}
\chi^2_{BAO}(p_s)=X^tV^{-1}X,\label{eq:chi2BAO}
\end{equation}
where $X$ is a column vector formed from the values of theory minus
the corresponding observational data, with
\begin{eqnarray}
&&X= \left(
\begin{array}{c}
 \frac{r_s(z_d)}{D_V(0.2)}-0.1905 \\
 \frac{r_s(z_d)}{D_V(0.35)}-0.1097
\end{array}
\right),
\end{eqnarray}
and $X^t$ denotes its transpose.

\subsection{Cosmic Microwave Background constraints}

The CMB shift parameter $R$ is provided by \cite{ref:Bond1997}
\begin{equation}
R(z_{\ast})=\frac{\sqrt{\Omega_m
H^2_0}}{\sqrt{|\Omega_k|}}\mathrm{sinn}[\sqrt{|\Omega_k|}\int_0^{z{_\ast}}\frac{dz'}{H(z')}],
\end{equation}
here, the redshift $z_{\ast}$ (the decoupling epoch of photons) is
obtained by using the fitting function \cite{Hu:1995uz}
\begin{equation}
z_{\ast}=1048\left[1+0.00124(\Omega_bh^2)^{-0.738}\right]\left[1+g_1(\Omega_m
h^2)^{g_2}\right],\nonumber
\end{equation}
where the functions $g_1$ and $g_2$ read
\begin{eqnarray}
g_1&=&0.0783(\Omega_bh^2)^{-0.238}\left(1+ 39.5(\Omega_bh^2)^{0.763}\right)^{-1},\nonumber \\
g_2&=&0.560\left(1+ 21.1(\Omega_bh^2)^{1.81}\right)^{-1}.\nonumber
\end{eqnarray}
In additional, the acoustic scale is related to the distance ratio
and is expressed as
\begin{eqnarray}
&&l_A=\frac{\pi}{r_s(z_{\ast})}\frac{c}{\sqrt{|\Omega_k|}}\mathrm{sinn}[\sqrt{|\Omega_k|}\int_0^{z_\ast}\frac{dz'}{H(z')}].
\end{eqnarray}

\begin{table}[htbp]
\begin{center}
\begin{tabular}{c|lll}
\hline\hline
  &\ $\mathrm{5-year}$ $\mathrm{ML}$ &\ $\mathrm{5-year}$ $\mathrm{mean}$ &\ $\mathrm{error}$, $\mathrm{\sigma}$ \\ \hline
 $l_A(z_\ast)$ &\ $302.10$ &\ $302.45$ &\ $0.86$ \\ \hline
 $R(z_\ast)$ &\ $1.710$ &\ $1.721$ &\ $0.019$ \\ \hline
 $z_{\ast}$  &\ $1090.04$ &\ $1091.13$ &\ $0.93$ \\
\hline
\end{tabular}
\end{center}
\caption{\label{cmbdata} The observational $l_A, R, z_{\ast}$
data~\cite{ref:Komatsu2008}.}
\end{table}

Using the data of $l_A, R, z_{\ast}$ in \cite{ref:Komatsu2008},
which are listed in Table \ref{cmbdata}, and their covariance matrix
of $[l_A(z_\ast), R(z_\ast), z_{\ast}]$ referring to
\cite{ref:Komatsu2008}:
\begin{eqnarray}
&&C^{-1}= \left(
\begin{array}{ccc}
 1.800 & 27.968 & -1.103\\
 27.968 & 5667.577 & -92.263\\
 -1.103 & -92.263 & 2.923
\end{array}
\right),
\end{eqnarray}
we can calculate the likelihood $L$ as $\chi^2_{CMB}=-2\ln L$:
\begin{eqnarray}
&&\chi^2_{CMB}=\bigtriangleup d_i[C^{-1}(d_i,d_j)][\bigtriangleup
d_i]^t,
\end{eqnarray}
where $\bigtriangleup d_i=d_i-d_i^{data}$ is a row vector, and
$d_i=(l_A, R, z_{\ast})$.

\subsection{OHD}

The observational Hubble data are based on differential ages of the
galaxies \cite{ref:JL2002}. In \cite{ref:JVS2003}, Jimenez {\it et
al.} obtained an independent estimate for the Hubble parameter using
the method developed in \cite{ref:JL2002}, and used it to constrain
the EOS of dark energy. The Hubble parameter depending on the
differential ages as a function of redshift $z$ can be written in
the form of
\begin{equation}
H(z)=-\frac{1}{1+z}\frac{dz}{dt}.
\end{equation}
So, once $dz/dt$ is known, $H(z)$ is obtained directly
\cite{ref:SVJ2005}. By using the differential ages of
passively-evolving galaxies from the Gemini Deep Deep Survey (GDDS)
\cite{ref:GDDS} and archival data
\cite{ref:archive1,ref:archive2,ref:archive3,ref:archive4,ref:archive5,ref:archive6},
Simon {\it et al.} obtained $H(z)$ in the range of $0.1\lesssim z
\lesssim 1.8$ \cite{ref:SVJ2005}. In \cite{ref:0907}, Stern {\it et
al.} used the new data of the differential ages of
passively-evolving galaxies at $0.35<z<1$ from Keck observations,
SPICES survey and VVDS survey. The twelve observational Hubble data
from \cite{ref:0905,ref:0907,ref:SVJ2005} are list in Table
\ref{Hubbledata}. Here, we use the value of Hubble constant
$H_0=74.2\pm3.6 {\rm km ~s}^{-1} {\rm Mpc}^{-1}$, which is obtained
by observing 240 long-period Cepheids in \cite{ref:0905}. As pointed
out in \cite{ref:0905}, the systematic uncertainties have been
greatly reduced by the unprecedented homogeneity in the periods and
metallicity of these Cepheids. For all Cepheids, the same instrument
and filters are used to reduce the systematic uncertainty related to
flux calibration.
\begin{table}[htbp]
\begin{center}
\begin{tabular}{c|llllllllllll}
\hline\hline
 $z$ &\ 0 & 0.1 & 0.17 & 0.27 & 0.4 & 0.48 & 0.88 & 0.9 & 1.30 & 1.43 & 1.53 & 1.75  \\ \hline
 $H(z)\ ({\rm km~s^{-1}\,Mpc^{-1})}$ &\ 74.2 & 69 & 83 & 77 & 95 & 97 & 90 & 117 & 168 & 177 & 140 & 202  \\ \hline
 $1 \sigma$ uncertainty &\ $\pm 3.6$ & $\pm 12$ & $\pm 8$ & $\pm 14$ & $\pm 17$ & $\pm 60$ & $\pm 40$
 & $\pm 23$ & $\pm 17$ & $\pm 18$ & $\pm 14$ & $\pm 40$ \\
\hline
\end{tabular}
\end{center}
\caption{\label{Hubbledata} The observational $H(z)$
data~\cite{ref:0905,ref:0907}.}
\end{table}
In addition, in \cite{ref:0807}, the authors took the BAO scale as a
standard ruler in the radial direction, called "Peak Method",
obtaining three more additional data: $H(z=0.24)=79.69\pm2.32,
H(z=0.34)=83.8\pm2.96,$ and $H(z=0.43)=86.45\pm3.27$, which are
model and scale independent. Here, we just consider the statistical
errors.

 The best fit values of the model parameters from
observational Hubble data \cite{ref:SVJ2005} are determined by
minimizing
\begin{equation}
\chi_{Hub}^2(p_s)=\sum_{i=1}^{15} \frac{[H_{th}(p_s;z_i)-H_{
obs}(z_i)]^2}{\sigma^2(z_i)},\label{eq:chi2H}
\end{equation}
where $p_s$ denotes the parameters contained in the model, $H_{th}$
is the predicted value for the Hubble parameter, $H_{obs}$ is the
observed value, $\sigma(z_i)$ is the standard deviation measurement
uncertainty, and the summation is over the $15$ observational Hubble
data points at redshifts $z_i$.

\end{document}